\documentclass[aps,pra,twocolumn,superscriptaddress, amsmath,amssymb,10pt]{revtex4-2}
\usepackage[colorlinks=true,linkcolor=blue,urlcolor=blue,citecolor=blue,pdfusetitle]{hyperref}
\usepackage{graphicx}
\usepackage{braket}
\usepackage{color}
\usepackage{orcidlink}

\begin{document}
\title{Coordinate space representation of a one-dimensional odd-parity pseudopotential}

\author{Marc Camus-Sais~\orcidlink{0009-0004-0512-3752}}
\email{marc.camus.sais@gmail.com}
\affiliation{Departament de F{\'i}sica Qu{\`a}ntica i Astrof{\'i}sica, Facultat de F{\'i}sica, Universitat de Barcelona, E-08028 Barcelona, Spain.}
\affiliation{Institut für Quantenoptik und Quanteninformation, Österreichische Akademie der Wissenschaften, Innsbruck 6020, Austria.}

\author{Abel Rojo-Francàs~\orcidlink{0000-0002-0567-7139}}
\email{abel.rojo@oist.jp}
\affiliation{Departament de F{\'i}sica Qu{\`a}ntica i Astrof{\'i}sica, Facultat de F{\'i}sica, Universitat de Barcelona, E-08028 Barcelona, Spain.}
\affiliation{Institut de Ci{\`e}ncies del Cosmos, Universitat de Barcelona, ICCUB, Mart{\'i} i Franqu{\`e}s 1, E-08028 Barcelona, Spain.}
\affiliation{Quantum Systems Unit, Okinawa Institute of Science and Technology Graduate University, Onna, Okinawa 904-0495, Japan.}

\author{Grigori E. Astrakharchik~\orcidlink{0000-0003-0394-8094}}
\email{grigori.astrakharchik@upc.edu}
\affiliation{Departament de F\'isica, Universitat Polit\`ecnica de Catalunya, 
Campus Nord B4-B5, E-08034 Barcelona, Spain.}

\author{Bruno Juliá-Díaz~\orcidlink{0000-0002-0145-6734}}
\email{bruno@fqa.ub.edu}
\affiliation{Departament de F{\'i}sica Qu{\`a}ntica i Astrof{\'i}sica, Facultat de F{\'i}sica, Universitat de Barcelona, E-08028 Barcelona, Spain.}
\affiliation{Institut de Ci{\`e}ncies del Cosmos, Universitat de Barcelona, ICCUB, Mart{\'i} i Franqu{\`e}s 1, E-08028 Barcelona, Spain.}

\begin{abstract}
We propose a discrete-space representation of a one-dimensional zero-range odd-parity pseudopotential. The proposed representation is validated by applying it to the analytically solvable case of two fermions in a harmonic trap and successfully recovering the exact energy spectrum and eigenfunctions. 
Furthermore, we use the square-well and modified Pöschl--Teller potentials as finite-range representations of the odd-parity interaction and study their convergence to the contact interaction when the range tends to zero. 
Finally, we perform natural orbital analysis and compute the eigenvalues of the one-body density matrix for different particle numbers, examining their dependence on the one-dimensional scattering length and identifying distinct physical regimes.
\end{abstract}

\maketitle

\section{Introduction}
The use of ultracold gases as an experimental platform enables precise control over interactions between atoms, as well as fine-tuning of system parameters, including geometry, dimensionality, and particle number~\cite{ultracold_gases, simulators_lewensein, matter2, matter1}. For instance, Feshbach resonances can be used to tune the strength and even the sign of interatomic interactions with high accuracy, enabling the exploration of regimes ranging from weakly interacting Bose-Einstein condensates to strongly correlated Fermi gases~\cite{feshbach1, feshbach2}. Optical lattices can be engineered to mimic crystalline solids with adjustable lattice spacing and dimensionality~\cite{optical_lattice1,optical_lattice2}, facilitating the study of quantum phase transitions such as the superfluid-to-Mott-insulator transition~\cite{super-mott}. Furthermore, by tailoring spin-dependent interactions and synthetic gauge fields, ultracold gases have been employed to realize exotic phases of matter, including topological superfluids and quantum spin liquids~\cite{spin-liquids}. This high degree of control makes them an ideal testbed for simulating complex quantum many-body systems that are otherwise challenging to access in conventional condensed matter settings.

Quantum fluctuations are enhanced as dimensionality is reduced, being stronger in $2D$ and even more pronounced in $1D$. One-dimensional systems, in particular, are often easier to handle numerically and may even admit exact solutions~\cite{1dimensional1, 1dimensional2}. Significant advances in trapping and cooling techniques~\cite{cooling&trapping1, cooling&trapping2} have enabled the confinement of cold atomic gases in a one-dimensional regime, achieved using waveguides with tight transverse confinement~\cite{waveguide1,waveguide2}. The quasi-$1D$ regime is realized when the chemical potential and the thermal energy are smaller than the energy level spacing of the transverse oscillator~\cite{1D_conditions}. Under these conditions, the longitudinal motion prevails over other directions, and the system can effectively be considered a one-dimensional harmonic oscillator.

In ultracold quantum gases, interactions are generally dominated by $s$‑wave scattering~\cite{scattering}. However, in systems of identical fermions with spin fully polarized, the Pauli exclusion principle forces the spatial wave function to be antisymmetric, which completely forbids $s$‑wave contact interactions. Consequently, the leading interaction channel becomes $p$‑wave scattering. This has been observed and explored both experimentally and theoretically, for instance, in measurements of a $p$-wave Feshbach resonance in ultracold $^{40}K$ gases~\cite{Regal2003} and in studies of $p$-wave resonant threshold behavior~\cite{DeMarco1999}. Subsequent work has investigated $p$-wave collisions in low-dimensional geometries~\cite{Günter2005}, and more recent theoretical and quantum Monte Carlo studies have characterized the roles of the $p$-wave scattering volume and effective range in spin-polarized Fermi systems~\cite{Bertaina2023}. Motivated by this body of work, we analyze properties of a system of $N$ spin-aligned fermions in a $1D$ harmonic trap, where the spin state is frozen in the fully polarized configuration $\uparrow_1\uparrow_2\ldots\uparrow_N$, and $p$‑wave scattering is the only allowed short‑range contact interaction.

In one-dimensional systems, the notion of a $p$-wave interaction must be reformulated, since the absence of angular degrees of freedom precludes the usual partial-wave expansion familiar from three dimensions. Instead, what is conventionally referred to as a $p$-wave interaction in $1D$ is more precisely captured by an odd-parity pseudopotential, which enforces the appropriate symmetry of the relative wave function under particle exchange. This odd-parity contact interaction reproduces the low-energy scattering properties associated with the $3D$ $p$-wave channel when the system is strongly transversely confined, giving rise to effective $1D$ scattering parameters such as the scattering length and effective range. 

The structure of this work is as follows. In Sec.~\ref{sec:model}, we introduce the Hamiltonian of the system and show two limits for which analytical solutions are known. In Sec.~\ref{sec:discrete}, we derive a representation of the odd-parity interaction in discrete space, which reproduces $p$-wave scattering in the zero-range limit. In Sec.~\ref{sec:continuous}, we present two possible representations of the interaction in continuous space using two different potentials: a square-well and the modified Pöschl--Teller potential. This framework allows us to model the odd-parity interaction as a function of the interaction range. In Sec.~\ref{sec:results}, we present the relative energy spectrum of a two-fermion system, using the different representations of the interaction derived in the previous sections. Moreover, we study the one-body density matrix for different scattering lengths and numbers of fermions. Finally, in Sec.~\ref{sec:conclusions}, we summarize and present the main conclusions of our work.

\section{Model}\label{sec:model}
Consider $N$ spin-aligned fermions confined in a $1D$ harmonic trap and interacting via an odd-parity zero-range pseudopotential. The Hamiltonian of the system is~\cite{mapping_olshanii}
\begin{equation}\label{eq:hamiltonian}
\hat{\mathcal{H}} = \sum_{i=1}^N-\frac{\hbar^2}{2m}\partial_{x_i}^2 + \frac{1}{2}m\omega^2x_i^2-g_\text{F}\sum_{j>k}^N\delta'(x_{kj})\hat{\partial}_{kj},
\end{equation}
where $\delta'(x_{kj}) \equiv \partial_{x_{kj}}[\delta(x_{kj})]$ being $x_{kj}\equiv x_k - x_j$, and $\hat{\partial}_{kj}$ is the regularized operator
\begin{equation}
\hat{\partial}_{kj}\psi = \frac{1}{2}\left(\partial_{x_k}\psi\big|_{x_k = x_j^+}-\partial_{x_j}\psi\big|_{x_k = x_j^-}\right).
\end{equation}
The one-dimensional scattering length, $a_{1D}$, is defined as the position of the node, closest to $x_{kj}=0$, obtained by extrapolating the asymptotic linear behavior of the two-body scattering wave function
\begin{equation}\label{Eq:a1D}
\psi(x_{kj}) \propto x_{kj} - a_{1D},\quad x_{kj}\to\infty.
\end{equation}
The zero-range interaction potential, $-g_\text{F}\delta'(x_{kj})\hat{\partial}_{kj}$, is characterized by the coupling constant related to the scattering length as $g_\text{F} = 2a_{1D}\hbar^2/m$. Thus, for a zero-range potential, the scattering length is related to the Bethe-Peierls condition for the logarithmic derivative of the wave function at the origin,
\begin{equation}\label{Eq:boundary_condition}
\psi'(0^+)/\psi(0^+) = - 1/a_{1D}.
\end{equation}
The use of a simple delta-function contact potential for $p$-wave fermions is problematic, as it cannot reproduce the correct boundary condition Eq.~(\ref{Eq:boundary_condition}) while preserving antisymmetry under particle exchange. Since a $\delta$-potential acts only when two fermions occupy the same position, it has no effect on antisymmetric wave functions that vanish at contact.
Consequently, odd-parity interactions cannot be directly modeled using such simple contact potentials, including those typically employed in exact-diagonalization methods. A proper description requires either derivative (slope-dependent) zero-range potentials or the appropriate finite-range limit, which might be more involved to implement numerically.

Analytic solutions for a general number of particles are known in two limits, the non-interacting case and the infinitely interacting limit. In the latter, the system is solvable by mapping the original Hamiltonian to that of an ideal Bose gas~\cite{FTG_gas}. The ground state energy of infinitely interacting fermions is equal to that of non-interacting bosons. In this limit, the system is known as the fermionic Tonks-Girardeau (FTG) gas. This mapping is analogous to the well-known correspondence between an impenetrable Bose gas and an ideal Fermi gas~\cite{TG_gas, tonks_gas}, where the ground state energy of non-interacting fermions matches that of infinitely repulsive bosons. Outside these limits, no analytic solution is known for $N>2$, and the problem must therefore be addressed numerically. 

\section{Discrete representation}\label{sec:discrete}

In this section, we propose an explicit discrete-space representation of the odd-parity interaction in the zero-range limit. In this framework, the $s$-wave interaction, $g_\text{B}\delta(x_{kj})$, is typically represented by a potential of depth $g_\text{B}/\Delta x$ at mesh points where $x_i = x_j$, with $\Delta x$ denoting the spatial discretization step. We extend this concept by introducing a potential that accurately captures the physics of the odd-parity interaction.

To derive the discrete representation, we take advantage of the existence of an exact solution to a fermionic problem in the continuous representation and invert the stationary Schrödinger equation to express the interaction potential $V$ in terms of the second derivative of the eigenfunction $\Psi(x_1,\ldots,x_N)$,
\begin{equation}\label{eq:v_int}
\begin{aligned}
V(x_1,\ldots,x_N) =& E + \frac{\hbar^2}{2m}\frac{\sum_{i = 1}^N\partial_{x_i}^2\Psi(x_1,\ldots,x_N)}{\Psi(x_1,\ldots,x_N)}\\
    &-U(x_1, \ldots,x_N),
\end{aligned}
\end{equation}
where $U(x_1,\ldots,x_N)$ denotes the external potential. Solving this equation requires prior knowledge of the eigenenergy $E$ and eigenfunction $\Psi$.

We consider the system of two spin-aligned fermions in free space with odd-parity zero-range interaction, for which the solutions are known analytically. 
The Hamiltonian of the system is
\begin{equation}
\hat{\mathcal{H}} = -\frac{\hbar^2}{2m}\partial_{x_1}^2-\frac{\hbar^2}{2m}\partial_{x_2}^2 -g_\text{F}\delta'(x_1-x_2)\hat{\partial}_{12}.
\end{equation} 
This Hamiltonian is separable in the center of mass (CM) coordinate $X$ and the relative position between the two particles $x$. The CM part has only the kinetic term leading to plane-wave solutions. For $g_\text{F}>0$, the relative Hamiltonian has a bound state with wave function~\cite{free_fermions}
\begin{equation}\label{eq:psi_rel_free}
\psi_\text{rel}(x) = \sqrt{\frac{\hbar^2}{m g_\text{F}}}\text{sgn}(x) e^{-2|x|\hbar^2/(m g_\text{F})},
\end{equation}
with energy $E_\text{rel} = -4\hbar ^6/(m^3g_\text{F}^2)$. This solution is used to determine $V$ from Eq.~(\ref{eq:v_int}).

To perform the calculation numerically in position space, we discretize it. The choice of mesh directly determines the shape of the potential. At $x = 0$, the sign function is equal to $\text{sgn}(x=0)=0$, and Eq.~(\ref{eq:v_int}) gives $V(x = 0) \to \infty$, introducing a divergent Hamiltonian term that prevents numerical diagonalization. To avoid this issue, the mesh is defined as $\{-L, -L +\Delta x, \ldots, -\Delta x, \Delta x,\ldots , L-\Delta x, L\}$, explicitly excluding $x=0$. Within this grid, the discrete representation is non-zero only at the points $x = \pm\Delta x$, with height equal to
\begin{equation}\label{Eq:potential}
\begin{aligned}
V(\pm\Delta x) = & \frac{\hbar^2}{m}\frac{\psi_\text{rel}(0)-2\psi_\text{rel}(-\Delta x) + \psi_\text{rel}(-2\Delta x)}{\Delta x^2\psi_\text{rel}(-\Delta x)}\\
&+E_\text{rel} =\frac{\hbar^2}{m\Delta x^2}\left( e^{-\gamma}-2-\gamma^2\right), 
\end{aligned}
\end{equation}
where $\psi_\text{rel}(0) \equiv 0$, and $\gamma\equiv2\Delta x\hbar^2/(m g_\text{F})$. Figure~\ref{fig:potentials} shows the resulting potential as a function of the interparticle separation. Note that $x=0$ is absent in the mesh.

In summary, we have introduced an explicit discrete space representation of the odd-parity interaction valid in the zero-range limit ($R\to0$)
\begin{equation}
V(x_1,\ldots)=\sum_{i <j}^N \frac{\hbar^2}{m\Delta x^2}\left( e^{-\gamma}-2 -\gamma^2\right)\delta_{x_i,x_j\pm\Delta x},
\label{Eq:V:discrete}
\end{equation}
where $\delta_{x_i,x_j\pm\Delta x}$ is the Kronecker delta. This interaction is non-zero only at $x_i=x_j\pm\Delta x$. Moreover, this formulation is derived assuming that the discrete second derivative is computed as
\begin{equation}
\partial^2_x\psi(x) = \frac{\psi(x+\Delta x)-2\psi(x)+\psi(x-\Delta x)}{\Delta x^2},
\end{equation}
which must be used consistently to ensure the correctness of the potential.

The Hamiltonian with potential Eq.~(\ref{Eq:potential}) is invariant under the exchange of any two particles, leading to $N!$ degenerate energy levels. To obtain the fermionic states, we introduce a symmetry-breaking term in the Hamiltonian that assigns an energy penalty to eigenstates lacking the required antisymmetry, thereby restricting the low-energy manifold to purely fermionic wave functions. This additional term appears in the off-diagonal elements of the Hamiltonian as
\begin{equation}
\begin{split}
V_\mathrm{odd}\left[\psi(x_1,\ldots) \right]
&= K \sum_{P \in \mathcal{S}_N} (-1)^{\epsilon(P)} \psi(x_{P_1},\ldots),
\end{split}
\end{equation}
where $\mathcal{S}_N$ denotes the Symmetric group, and $(-1)^{\epsilon(P)}$ reflects the parity of the permutation $P$. If $P$ can be written as an even number of transpositions, it evaluates to $+1$; if an odd number of transpositions is required, it evaluates to $-1$. This term contributes an energy of $N!K$ to the fermionic states. By choosing $|K|\gg1$ and $K <0$, all the fermionic solutions are shifted to the low-energy sector. As a result, the physically relevant eigenvalues can be obtained simply by computing the smallest eigenvalues of the full Hamiltonian.

Although Eq.~(\ref{Eq:V:discrete}) is originally derived for two fermions in free space with an odd-parity interaction and a positive scattering length, it yields accurate results for harmonically trapped systems for both positive and negative scattering lengths. 

\section{Continuous representation}\label{sec:continuous}

In this section, we study two continuous-space representations of the odd-parity interaction using two different potentials: a square-well and the modified Pöschl--Teller potential~\cite{posch_teller}. We focus on the case of two spin-polarized fermions in free space and determine how the potential depths relate to the scattering length and the interaction range, $R$.

\subsection{Square-well potential}
We work in the CM reference frame. The potential as a function of the relative distance between particles, $x$, is 
\begin{equation}
    V_{SW}(x) = \left\{
	\begin{aligned}
	    - &V_0  &\quad &\text{if} &|x| < R_{SW} \\
		&0 &\quad &\text{if} &|x| \ge R_{SW}
	\end{aligned}
\right.\,, 
\end{equation}
where $R_{SW}$ is the range of the interaction and $V_0$ is positive. In Fig.~\ref{fig:potentials} we depict the form of the potential. At low energies, the relative-motion Schrödinger equation reads
\begin{equation}
        \left\{
	\begin{aligned}
	   &\partial^2_x\psi(x) + \kappa^2\psi(x) = 0  &\quad &\text{if}& |x| < R_{SW} \\
		 &\partial^2_x\psi(x) = 0 &\quad &\text{if}& |x| \geq R_{SW}
	\end{aligned}
    \right.\,,
\end{equation}
with $\kappa= \sqrt{m V_0/\hbar^2}$ being the characteristic wavenumber related to the depth of the interaction potential. Due to fermionic antisymmetry, the solution must have odd parity. The zero-energy scattering solution to the Schrödinger equation is
\begin{equation}
\left\{
\begin{aligned}
     \psi(x) &= N_2(x+a_{1D}) &\quad &\text{if} && x  \leq  -R_{SW}\\
     \psi(x) &= N_1\sin(\kappa x)    &\quad &\text{if} &|&x|  <  R_{SW} \\
     \psi(x) &= N_2(x-a_{1D}) &\quad &\text{if} && x \geq R_{SW}
\end{aligned}
\right.\,,
\end{equation}
where $a_{1D}$ is the one-dimensional scattering length. To relate the potential depth, the scattering length, and the interaction range, one enforces continuity of the logarithmic derivative of $\psi(x)$ at $x=R_{SW}$. This yields the relation between the scattering length and parameters of the interaction potential~\cite{LandauLifshitzQM}
\begin{equation}\label{eq:square_well}
a_{1D} = R_{SW}\left[1 - \frac{\tan(\kappa R_{SW})}{\kappa R_{SW}}\right].
\end{equation}
We assume that $\kappa R_{SW}<4.493$, limiting the maximum value of the potential depth, and ensuring that $\psi(x)$ has only a single node at $x=0$ inside the well. The effective range for the square-well potential is
\begin{equation}\label{eq:square_well:reff}
r_\text{eff} = R_{SW}\left[1 - \frac{R_{SW}^2}{3a_{1D}^2} -  \frac{1}{\kappa^2 a_{1D}R_{SW}}\right].
\end{equation}

\subsection{Modified Pöschl--Teller potential}

\begin{figure}[t]
\centering
\includegraphics[width=\linewidth]{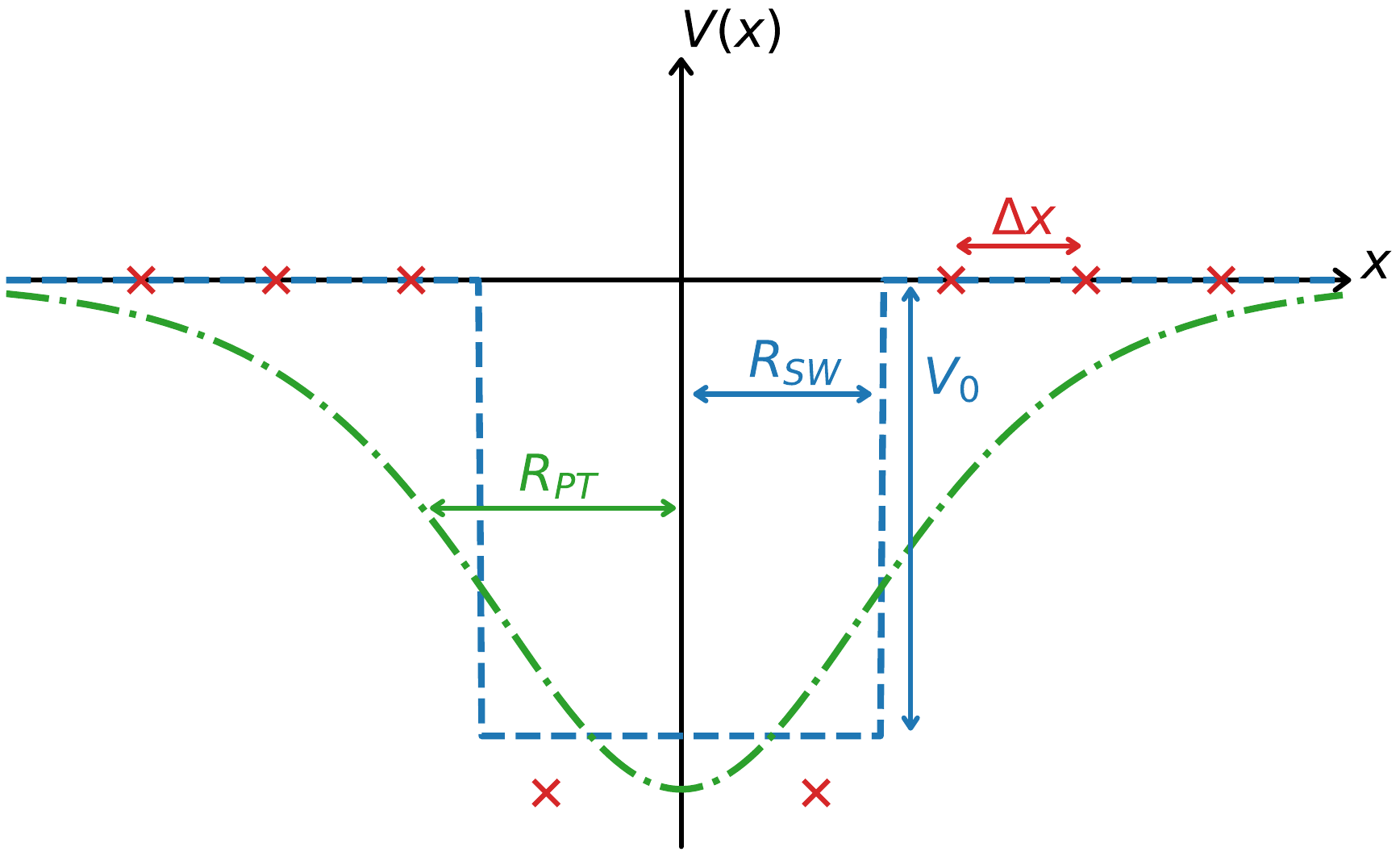}
\caption{Discrete and continuous representations of the odd-parity interaction as a function of the relative coordinate between fermions, $x$. The discrete representation is shown as red crosses, with no specific point defined at $x=0$. The square-well potential, $V_{SW}$, is depicted by a dashed blue line while the modified Pöschl--Teller potential, $V_{PT}$, is shown by a dash-dotted green line.}
\label{fig:potentials}
\end{figure}

The Modified Pöschl--Teller potential is defined as
\begin{equation}
V_{PT}(x) = -\frac{\hbar^2}{m R_{PT}^2}\frac{\lambda(\lambda-1)}{\cosh^2(x/R_{PT})},
\label{Eq:PoshlTeller}
\end{equation}
where $\lambda >1$ is a dimensionless parameter defining the interaction strength. 
While formally potential Eq.~(\ref{Eq:PoshlTeller}) has an infinite range, the distance $R_{PT}$ effectively defines its physical interaction range. Figure~\ref{fig:potentials} illustrates the shape of the modified Pöschl--Teller potential. For positive energies and at large positive distance, the asymptotic behavior of the odd-parity eigenfunctions is given by~\cite{flugge1971practical}
\begin{equation}\label{eq:asym}
\begin{aligned}
    \psi(x)\to\frac{\sqrt{\pi}}{2}\Bigg\{&\frac{\Gamma\left(-ikR_{PT}\right)e^{ik\ln(2)R_{PT}}e^{-ik|x|}}{\Gamma\left(\frac{\lambda+1}{2}-\frac{ikR_{PT}}{2}\right)\Gamma\left(1-\frac{\lambda}{2}-\frac{ikR_{PT}}{2}\right)}\\
    &+\text{c.c.}\Bigg\},
\end{aligned}
\end{equation}
where $k^2 = m E/\hbar^2$. Expanding Eq.~(\ref{eq:asym}) for $k\to0$, a linear series decomposition allows one to locate the node of the wave function and thereby obtain the $1D$ scattering length
\begin{equation}\label{eq:pt_amplitude_potential}
    \frac{a_{1D}}{R_{PT}} = \frac{\pi}{2}\text{cotan}\frac{\pi\lambda}{2}+\gamma+\Psi(\lambda),
\end{equation}
where $\gamma = 0.5772\ldots$ is Euler’s constant and $\Psi(\lambda)$ denotes the Digamma function. 

\subsection{Soliton wave function}

In the limit of a small and positive scattering length, the ground state solution in the absence of harmonic trapping corresponds to a bright soliton, for which McGuire found an explicit expression for the many-body wave function~\cite{free_fermions}.
In the presence of harmonic trapping, a related wave function was proposed in Ref.~\cite{Beau2020} and reads as
\begin{equation}\label{Eq:wf:a>0}
    \Psi(x_1,\ldots,x_N) = \mathcal{N}\prod_{i<j}^Ne^{-|x_{ij}|/a_{1D}}\prod_i^Ne^{-x_i^2/(2a_\mathrm{ho}^2)},
\end{equation}
with $a_{1D}>0$.
In the absence of harmonic confinement ($a_\mathrm{ho}\to\infty$), wave function Eq.~(\ref{Eq:wf:a>0}) coincides with the wave function of the McGuire soliton~\cite{free_fermions}.
Wave function Eq.~(\ref{Eq:wf:a>0}) is an exact solution of Hamiltonian Eq.~(\ref{eq:hamiltonian}) in the presence of an additional linear interaction $\sigma |x|$ with $\sigma = -m\omega g /\hbar$. 
In the limit of strong attraction, $a_{1D}\to +0$, the soliton size is of the order of $\xi \propto a_{1D}/N$, so that the linear potential $\sigma |x|$ is sampled in the vicinity of its zero value, so its presence does not affect much the solution. In other words, the linear potential can be neglected in the $a_{1D}\to +0$ limit and wave function Eq.~(\ref{Eq:wf:a>0}) can be used for describing system properties in $a_{1D}\to 0$ limit.

For negative scattering lengths, no bound state is formed in the two-body scattering problem, so instead of $e^{-|x_{ij}|/a_{1D}}$ terms we consider $|x_{ij}|-a_{1D}$ Jastrow terms, which correspond to zero-energy two-body scattering solution, resulting in the following wave function, 
\begin{equation}
\Psi_\text{B}(x_1,\ldots,x_N) = \prod\limits_{i<j}^N (|x_{ij}|-a_{1D})\prod\limits_{i=1}^N e^{-x_i^2/(2a_\mathrm{ho}^2)},
\label{Eq:wf:a<0}
\end{equation}
with $a_{1D}<0$.

The wave functions Eqs.~(\ref{Eq:wf:a>0}-\ref{Eq:wf:a<0}) are symmetric under exchange of two particles, and, thus, satisfy Bose-Einstein statistics. The fermionic symmetry can be imposed by multiplying the wave functions by the antisymmetric function, $\Psi_F = \mathcal{A}\Psi_B$, where

\begin{equation}
\mathcal{A}(x_1,\ldots,x_N)=\prod_{ i<j}^N\text{sgn}(x_i-x_j),
\label{Eq:wf:antisymmetrization}
\end{equation}
which corresponds to Girardeau mapping~\cite{TG_gas}. 
Defined in this way, the fermionic wave functions $\Psi_F$ change the sign, whenever positions of two particles are exchanged.

For a state for which the many-body wave function is known, Monte Carlo methods can be efficiently used to sample its properties, especially, when it can be written in a pair-product for, as in Eqs.~(\ref{Eq:wf:a<0}-\ref{Eq:wf:antisymmetrization}). 
We will use it for the calculation of the OBDM and its eigenvalues.

\section{Results}\label{sec:results}
In this section, we present the numerically computed energy spectrum obtained using both the discrete and continuous representations of the interaction for fermions confined in a harmonic potential. Furthermore, we calculate the eigenvalues of the one-body density matrix for different particle numbers and scattering lengths.

\subsection{Energy spectrum}
The energy spectrum of the Hamiltonian Eq.~(\ref{eq:hamiltonian}) for two fermions is known analytically~\cite{Busch1998,Blume}. The relative energy $E_\text{rel}$ is obtained by solving the equation
\begin{equation}\label{eq:energy_2_fermions}
\frac{a_{1D}}{\sqrt{2}a_\mathrm{ho}} = \frac{\Gamma\left(-E_\text{rel}/(2\hbar\omega)+1/4\right)}{2\Gamma\left(-E_\text{rel}/(2\hbar\omega)+3/4\right)},
\end{equation}
where $a_\mathrm{ho} = \sqrt{\hbar/(m\omega)}$ is the harmonic oscillator length. Both representations of the interaction introduced above allow us to compute the energy of this system. Although the continuous representation corresponds to a finite interaction range, as this range decreases, the energy must converge to Eq.~(\ref{eq:energy_2_fermions}).
\begin{figure}[t]
\centering
\includegraphics[width=\columnwidth]{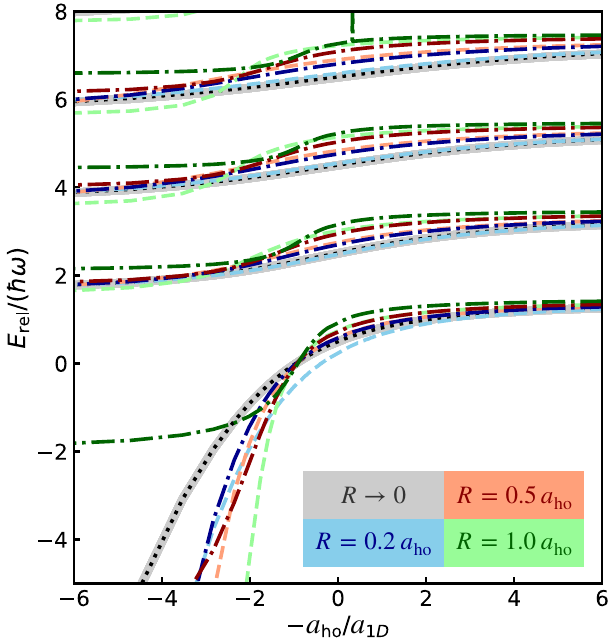} 
\caption{
Energy spectrum of the relative motion of two fermions in a $1D$ harmonic trap as a function of the interaction parameter. The solid gray line shows the odd solutions in the zero-range model, as given by Eq.~(\ref{eq:energy_2_fermions}), while the dotted black line corresponds to the energy obtained using the discrete representation of the potential Eq.~(\ref{Eq:V:discrete}) with a grid spacing of $\Delta x = 0.01\,a_\mathrm{ho}$ over the interval $x \in [-10, 10]\,a_\mathrm{ho}$. The two curves coincide perfectly.
Results for two models of finite-range continuous potentials are shown for potential range equal to $R=0.2,\,0.5,\,1.0\, a_\mathrm{ho}$ are shown in blue, red, and green, respectively. Dashed lines correspond to the calculations with the square-well and the dash-dotted lines to the modified Pöschl--Teller potentials.
}
\label{fig:spectrum}
\end{figure}

\begin{figure}[t]
\centering
\includegraphics[width=1\linewidth]{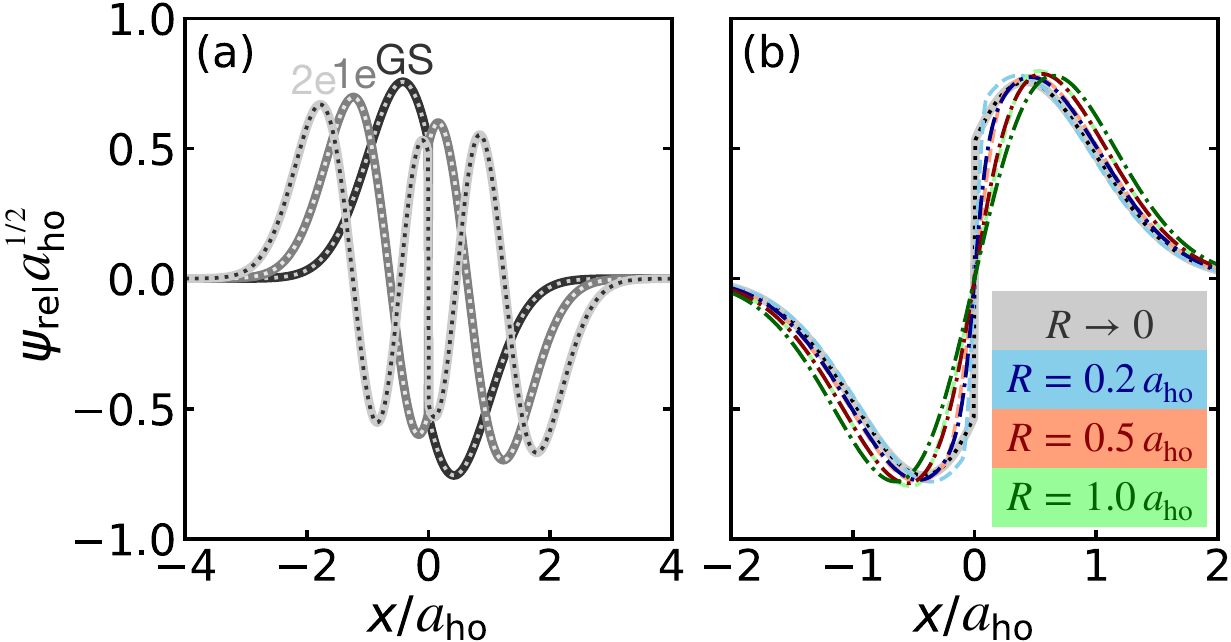} 
\caption{Relative wave functions of two fermions in a $1D$ harmonic trap. Panel (a) shows the three lowest relative wave functions for $a_{1D} = -a_\mathrm{ho}$, computed analytically and with the discrete representation. Panel (b) compares the ground state relative wave function for $a_{1D} = -a_\mathrm{ho}$ using the discrete and continuous representations.}
\label{fig:eigenfunctions}
\end{figure}
Figure~\ref{fig:spectrum} shows the energy spectrum in the zero-range limit. The analytical solution is shown with a solid gray line, while the numerical results obtained using the discrete representation are shown as a dotted black line. Both results are in complete agreement, demonstrating that the discrete representation, indeed, correctly reproduces the physics of the odd-parity interaction in the zero-range limit. Let us now examine how the energy depends on the interaction strength, expressed as the ratio of the harmonic oscillator length $a_\mathrm{ho}$ and the one-dimensional scattering length $a_{1D}$. In the non-interacting limit ($-a_\mathrm{ho}/a_{1D}\to +\infty$), the ground-state energy of the relative motion approaches $E_\text{rel}=(3/2)\hbar\omega$, corresponding to the occupation of the two lowest single-particle levels with energies $(1/2)\hbar\omega$ and $(3/2)\hbar\omega$, after subtracting the $(1/2)\hbar\omega$ CM contribution.
As the strength of the attractive interaction is increased, the energy decreases monotonically. 
In the Fermionic Tonks-Girardeau limit, the scattering length diverges and the interaction parameter passes through zero, $-a_\mathrm{ho}/a_{1D}\to 0$. 
At that point, the fermionic system has the same energy as two ideal bosons, so that $E_\text{rel} = \hbar\omega/2$. 
For positive values of the scattering length, in the continuum, a bound state enters at the threshold $1/a_{1D}=0$, with energy $E_b = -\hbar^2/(ma_{1D}^2)$. In the presence of harmonic confinement, however, no bound state exists at the value of the continuum threshold and the energy is positive. Still, for sufficiently small $a_{1D}>0$, the relative energy tends to the dimer energy $E_\text{rel}\to E_b$ as the bound state becomes more deeply bound and its reduced spatial extent makes it increasingly insensitive to the external confinement.

For the continuous finite-range interaction potentials, the energy approaches the zero-range limit as the interaction range is reduced, $R\to 0$. In dashed lines are shown the results using the square-well potential while in dash-dotted lines are for the modified Pöschl--Teller potential. For a given interaction range, the results obtained for the square-well potential are more similar to Eq.~(\ref{eq:energy_2_fermions}) than the ones obtained using the modified Pöschl--Teller potential.

The kinetic energy is directly related to the curvature of the wave function with larger curvature corresponding to a higher kinetic energy. The system energy $E_\text{rel}$ is increased as additional nodes are introduced in the relative wave function. However, due to its odd parity, each excitation adds two nodes. Consequently, the $n$-th excited relative wave function has $2n + 1$ nodes. The ground state for fermions has one node in its relative wave function, the first excited state has three nodes, the second has five, and so on. Figure~\ref{fig:eigenfunctions}(a) shows the three lowest eigenfunctions for $a_{1D} = -a_\mathrm{ho}$ computed both analytically and with the discrete representation. The results confirm this predicted behavior, with the numerically obtained wave functions for the ground and excited states closely matching the analytical solutions. Figure~\ref{fig:eigenfunctions}(b) compares the ground state wave function for $a_{1D}=-a_\mathrm{ho}$ obtained with the continuous and discrete representations. We observe that as the interaction range $R$ decreases, the eigenfunctions from the continuous representation progressively converge toward the discrete solution. 
\subsection{One-body density matrix and natural orbitals}
Since the energy spectrum of bosons with even-parity interactions is identical to that of fermions with odd-parity interactions, provided the interaction strengths are appropriately matched, both systems share the same energetic and diagonal properties. However, their off-diagonal properties are significantly different. It is therefore of particular interest to compute the one-body density matrix (OBDM) and perform a natural orbital analysis. The OBDM is defined as
\begin{equation}\label{eq:OBDM}
\rho^{(1)}(x,x') = N\int dx_2\ldots dx_N\Psi^*(x',x_\perp)\Psi(x,x_\perp), 
\end{equation}
where $x_\perp\equiv(x_2,\ldots,x_N)$. The OBDM quantifies the spatial loss of coherence in the system. 
We assume it is normalized to the number of particles $N$. The diagonal of the OBDM gives the particle density, $\rho(x)\equiv\rho^{(1)}(x,x)$.

\begin{figure}[t]
\centering
\includegraphics[width=1\linewidth]{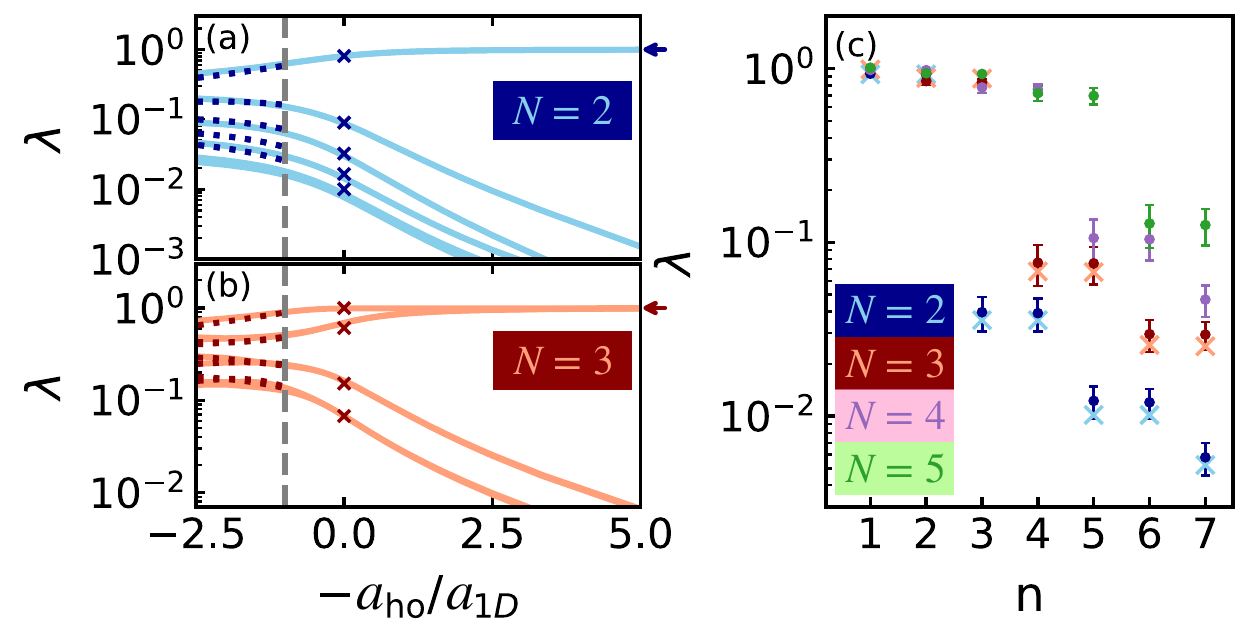} 
\caption{Eigenvalues of the OBDM for the ground state of a system of $N$ fermions confined in a harmonic trap. Panels (a) and (b) show the ten (for $N=2$) and seven (for $N=3$) largest eigenvalues, respectively, as a function of the interaction parameter. Numerical results using the discrete representation of the odd-parity potential are depicted as solid lines. The results obtained using Eq.~(\ref{Eq:wf:a>0}) are shown with dotted lines. 
The analytical results for the two limiting cases of zero and infinite scattering length~\cite{Francesc_OBDM_eigenvalues} are included for comparison. 
Specifically, the values in the FTG limit ($-a_{ho}/a_{1D}=0$) are represented by crosses, while the asymptotic values in the non-interacting limit ($-a_{ho}/a_{1D}\to\infty$) are indicated by arrows. In panel (a), all eigenvalues are doubly degenerate, whereas in panel (b), the largest eigenvalue is non-degenerate and the rest are doubly degenerate. Panel (c) shows the OBDM eigenvalues for $N=2,\,3,\,4,\,5$ at an intermediate scattering length, $a_{1D} = -a_\mathrm{ho}$. Cross markers represent results obtained using the ground state wave function obtained using the discrete representation of the odd-parity potential function Eq.~(\ref{Eq:V:discrete}) for $N=2,\,3$, while circular markers with error bars correspond to Monte Carlo results.}
\label{fig:eigenvalues}
\end{figure}

Figure~\ref{fig:eigenvalues}(a) shows the largest eigenvalues of the OBDM for the ground state of two fermions as a function of the interaction parameter. All the eigenvalues are doubly degenerate (i.e., there exist two orthogonal natural orbitals sharing the same occupation number). In the non-interacting limit, the two largest eigenvalues converge to unity, while the others approach zero. As the inverse of the scattering length goes from $-\infty$ to $0$, the two largest eigenvalues decrease and the remaining eigenvalues increase. In the FTG limit, $1/a_{1D}\to0$, the eigenvalues have the analytical form~\cite{Francesc_OBDM_eigenvalues,eigen_N2}
\begin{equation}
\lambda_n^{(2)} = \frac{8}{(\pi(2n-1))^2},
\end{equation}
where $n=1,2,\ldots$. If the inverse of the interaction strength is further increased, the two largest eigenvalues decrease to their minima while the others reach their maxima. In this regime, there is a strong attraction between the particles, so a dimer is formed. The relative wave function has a pronounced peak at $x_1 = x_2^\pm$, meaning that knowledge of the position of one particle fully determines the others. 
We see that the eigenvalues of the OBDM approach the ones found by using the soliton-like wave function Eq.~(\ref{Eq:wf:a>0}) properly antisymmetrized.  

Figure~\ref{fig:eigenvalues}(b) reports the largest OBDM eigenvalues for three fermions. All eigenvalues are doubly degenerate, except for the largest one, which is non-degenerate. In the non-interacting limit, $-a_{ho}/a_{1D}\to+\infty$, three eigenvalues converge to unity values while the others converge to zero. 
Indeed, this is what is expected for free fermions, each fermion occupies one single-particle state.
As the inverse of the scattering length varies from negative infinity to zero, the largest eigenvalue first decreases slightly and then increases, approaching unity. The second and third eigenvalues decrease, while the remaining ones increase, converging to the FTG limit. In this limit, the largest eigenvalue is $\lambda_0^{(3)} = 1$, and the remaining doubly degenerate eigenvalues are~\cite{Francesc_OBDM_eigenvalues} 
\begin{equation}
\lambda_n^{(3)} = \frac{24}{(2\pi n)^2}.
\end{equation}
As the inverse of the scattering length further increases, multiple eigenvalues contribute significantly. In this limit, the OBDM eigenvalues of Eq.~(\ref{Eq:wf:a>0}) are difficult to compute due to limited numerical resolution. We only show the convergence of the largest eigenvalue of the soliton model.  

Moreover, for the case of two and three fermions, we validate our numerical results using two independent methods. The first method exploits the discrete representation of the potential Eq.~(\ref{Eq:V:discrete}). For a small number of particles, it is feasible to discretize the coordinate space and diagonalize the Hamiltonian. However, since the size of the position space grows exponentially with the number of particles $N$, this approach becomes impractical for larger systems. To overcome this limitation, we employ the Diffusion Monte Carlo (DMC) method (see Ref.~\cite{SarsaBoronatCasulleras02} for a general reference). In our case, the fermionic wave function can be mapped to that of bosons, for which the complexity of the DMC algorithm scales as $N^2$, allowing us to efficiently access larger particle numbers. 
The guiding wave function used in Monte Carlo simulations is chosen by applying the antisymmetrization Eq.~(\ref{Eq:wf:antisymmetrization}) to the bosonic wave function Eq.~(\ref{Eq:wf:a<0}). The diffusion Monte Carlo method corrects the wave function and we use extrapolation to calculate the one-body density matrix.

Figure~\ref{fig:eigenvalues}(c) reports the seven largest OBDM eigenvalues for $N=2,\,3,\,4,\,5$ particles for a characteristic value of the repulsive interactions, $a_{1D} = -a_\mathrm{ho}$, shown with a dashed line in Figs~\ref{fig:eigenvalues}(a,b). For an even number of particles, all eigenvalues are doubly degenerate, while for an odd number of particles, all eigenvalues are doubly degenerate except for the largest one, which is non-degenerate. 
While bosons generally tend to occupy the lowest single-particle state, this is not the case for fermions. For the interaction strength $a_{1D} = -a_\mathrm{ho}$, we find that the lowest $n=1,\dots,N$ orbitals are strongly populated, with weights of order unity, whereas the occupation of higher orbitals is strongly suppressed. This behavior is characteristic of weakly interacting fermions. In contrast, in the fermionic Tonks–Girardeau regime ($-a_\mathrm{ho}/a_{1D}=0$), the occupations follow a qualitatively different distribution, scaling as $\propto 1/n^2$ for large $n$ and large particle numbers~\cite{Francesc_OBDM_eigenvalues}.

\section{Conclusions}\label{sec:conclusions}

In this work, we have studied a system of $N$ spin-aligned fermions with odd-parity interaction confined to a one-dimensional harmonic trap. We have introduced a discrete representation of the interaction, which has the correct zero-range limit and validated it by comparing its results with analytical solutions and with two continuous interaction models based on a square-well and a modified Pöschl--Teller potential. Finally, we have performed natural orbital analysis and have calculated the eigenvalues of the one-body density matrix as a function of the one-dimensional scattering length. 

Beyond the specific results presented here, this representation provides an alternative method to study systems with odd-parity interaction with the advantage that it eliminates the finite-range effects that arise when using the Pöschl--Teller or square-well potentials. Moreover, these methods can be used to numerically investigate fermion properties in more complex scenarios, such as in the presence of external fields.

\section{Acknowledgments}
This work has been funded by Grant PID2023-147475NB-I00 funded by MICIU/AEI/10.13039/501100011033 and FEDER, UE, by grants 
2021SGR01095 from Generalitat de Catalunya, and by 
Project CEX2019-000918-M of ICCUB (Unidad de Excelencia María 
de Maeztu).
G.E.A. acknowledges the support of the Spanish Ministry of Science and Innovation (MCIN/AEI/10.13039/501100011033, grant PID2023-147469NB-C21), the Generalitat de Catalunya (grant 2021 SGR 01411) and {Barcelona Supercomputing Center MareNostrum} ({FI-2025-1-0020}).
A.R-F. acknowledges additional funding from the Okinawa Institute of Science and Technology Graduate University.
\appendix
\section{Error analysis}
\begin{figure}[t]
\centering
\includegraphics[width=1\linewidth]{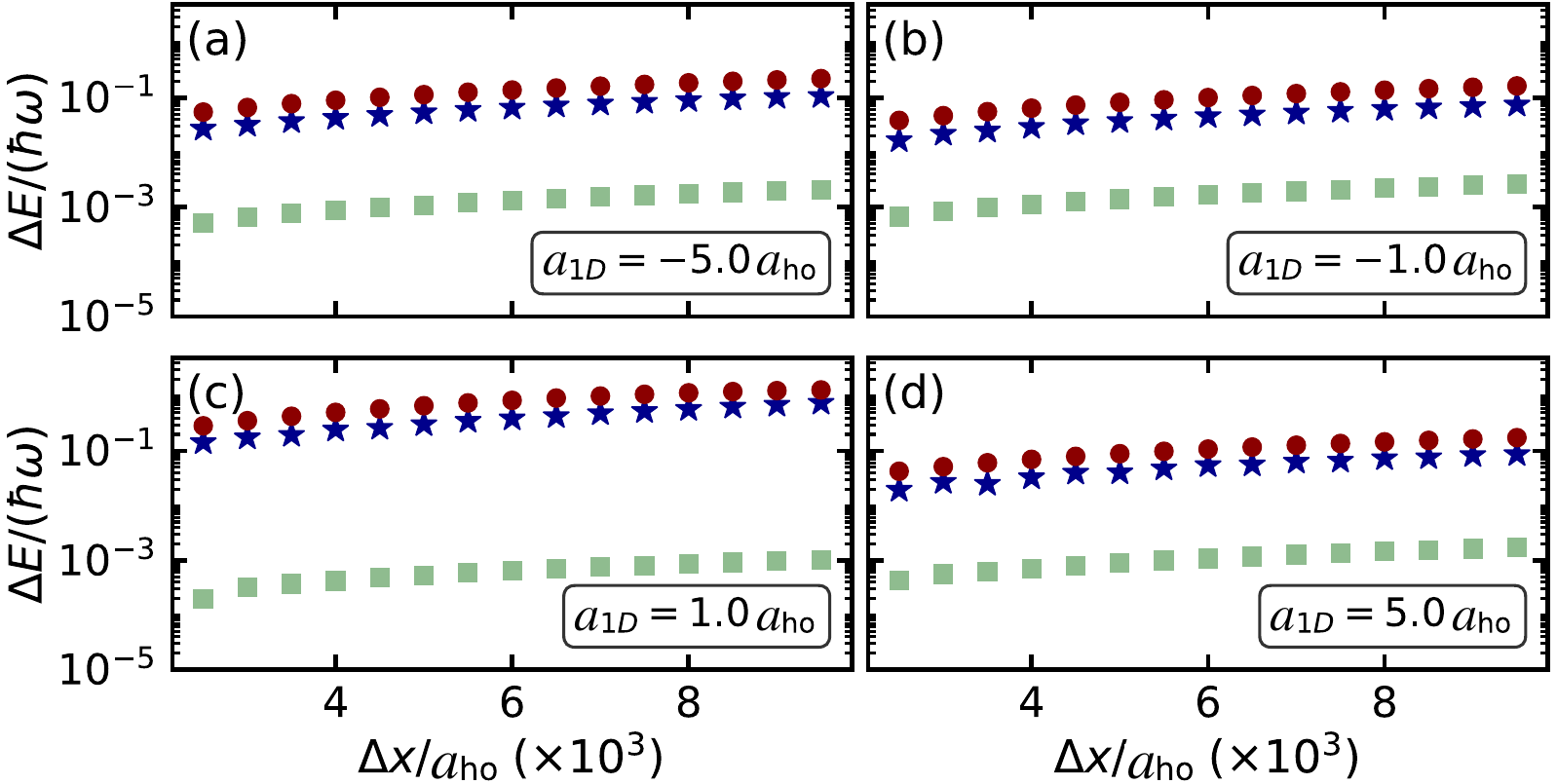} 
\caption{Absolute value of the energy difference $\Delta E \equiv|E_{\mathrm{Anl.}}-E_{\mathrm{Num}}|/(\hbar\omega)$ between the zero-range analytical result, Eq.~(\ref{eq:energy_2_fermions}), and the numerical results obtained using discrete and continuous representations, as a function of the grid spacing $\Delta x$ with $x \in [-10, 10]\,a_\mathrm{ho}$. Green squares, blue stars, and red circles correspond to results obtained using the discrete representation and the continuous representation with a square-well and a modified Pöschl–Teller potential, respectively. For the continuous representations, the interaction range is fixed to $R = 50\Delta x$. The convergence is analyzed for four scattering lengths: $a_{1D}/a_\mathrm{ho} = -5,\,-1,\,1,\,5$, shown in panels (a), (b), (c), and (d), respectively.}
\label{fig:error}
\end{figure}
In this section, we present an error analysis of the energy spectrum obtained using the discrete and continuous representations of the odd-parity pseudopotential. To this end, we consider the relative Hamiltonian of two interacting fermions in a harmonic trap,
\begin{equation}
    \hat{\mathcal{H}}_\text{rel} =-\frac{\hbar^2}{m}\frac{d^2}{dx^2}+\frac{1}{4}m\omega^2x^2 - g_\text{F}\delta'(x)\hat{\partial}_x.
\end{equation}
To assess the accuracy of the different representations, we first analyze the convergence of the numerical results using the discrete representation as the grid spacing $\Delta x$ is varied. As shown in Fig.~\ref{fig:error}, the ground state energy obtained from the numerical simulations approaches the analytical results, Eq.~(\ref{eq:energy_2_fermions}), as the grid spacing is reduced. We study the convergence for four different characteristic values of the scattering lengths. By performing a linear fit to the energy, $\Delta E/(\hbar\omega) = a\Delta x / a_\mathrm{ho}+ b$, we find that in the limit of small grid spacing, $\Delta x \to 0$, the discrepancy is as small as $\Delta E / (\hbar\omega)\sim10^{-5}$ for all considered cases. This result demonstrates the high accuracy of the discrete representation and its systematic convergence toward the zero-range solution.

While the discrete representation provides a systematic way to converge to the zero-range solution by refining the grid, alternative approaches model the interaction using finite-range continuous potentials. These representations, Eqs.~(\ref{eq:square_well},\ref{eq:pt_amplitude_potential}), depend explicitly on the interaction range, $R$. For a given grid spacing, the range of the interaction cannot be chosen arbitrarily small, since a sufficient number of grid points must lie within the potential region to properly capture the underlying physics. In our calculations, we therefore fix the interaction range to $R=50\Delta x$. As shown in Fig.~\ref{fig:error}, the energies obtained using these continuous representations also converge to the zero-range solution, Eq.~(\ref{eq:energy_2_fermions}) as the grid spacing is reduced, while maintaining the condition $R \gg \Delta x$. However, when comparing the three methods at the same computational cost (i.e., using the same grid spacing), the discrete representation is found to be significantly more accurate than the continuous ones. Specifically, we observe a ratio $\Delta E_\mathrm{cont} / \Delta E_\mathrm{disc} \sim 10^2$, indicating that continuous representations require substantially finer grids to achieve comparable accuracy. This increased computational demand limits the number of particles that can be efficiently simulated using finite-range continuous potentials.

For scattering lengths $a_{1D}/a_\mathrm{ho} = -5,\,-1,\,5$, the ratio $\Delta E_\mathrm{cont} / \Delta E_\mathrm{disc} \sim 10^2$, confirming the superior accuracy of the discrete representation in these regimes. In contrast, for $a_{1D}/a_\mathrm{ho} = 1 $, the difference between the discrete and continuous approaches becomes even more pronounced, with $\Delta E_\mathrm{cont} / \Delta E_\mathrm{disc} \sim 10^3$. In this regime, the relative wave function exhibits a large amplitude near the origin, indicating that the relevant physics is dominated by short-range behavior. As a consequence, the finite-range nature of the continuous representations makes it increasingly difficult to accurately capture the ground-state energy, in comparison with the discrete representation. This observation suggests that the discrete approach effectively suppresses finite-range effects inherent to square-well and modified Pöschl--Teller potentials. As the scattering length becomes smaller but remains positive, the wave function amplitude near the origin increases further, exacerbating this issue and leading to larger discrepancies for the continuous representations, as illustrated in Fig.~\ref{fig:spectrum}.

\bibliography{Biblio}

\end{document}